\documentclass[preprint]{ptephy_om}

\preprintnumber{RIKEN-iTHEMS-Report-25}

\usepackage{subcaption}

\usetikzlibrary{decorations.pathmorphing}
\tikzset{snake it/.style={decorate, decoration=snake}}

\numberwithin{equation}{section}

\usepackage{hyperref}

\title{Nonperturbative Formulation of Resonances in Quantum Mechanics Based on Exact WKB Method}

\author[1]{Okuto Morikawa\orcid{0000-0002-0044-4491}\thanks{okuto.morikawa@riken.jp}}
\affil[1]{Interdisciplinary Theoretical and Mathematical Sciences Program (iTHEMS),
RIKEN, Wako 351-0198, Japan}

\author[2]{Shoya Ogawa\orcid{0000-0003-0900-2486}\thanks{ogawa.shoya.615@m.kyushu-u.ac.jp}}
\affil[2]{Department of Physics, Kyushu University, 744 Motooka, Nishi-ku,
Fukuoka 819-0395, Japan}

\begin{document}
\begin{abstract}
We study quasi-stationary states in quantum mechanics using the exact Wentzel--Kramers--Brillouin (WKB) analysis as a nonperturbative framework. Whereas previous works focused mainly on stable systems, we explore unstable states such as resonances. As a concrete example, we analyze the inverted Rosen--Morse potential, which exhibits barrier resonance. This model allows exact solutions, enabling a direct comparison with exact WKB predictions. We provide a simple analytic picture of resonance and demonstrate consistency between exact and WKB-based results, extending the applicability of exact WKB analysis to nonpolynomial potentials.
\end{abstract}
\maketitle


\section{Introduction}
The nonperturbative analysis of quasi-stationary states in quantum mechanics (QM) has long been recognized as an important problem.
Recent developments in resurgence theory, particularly in the exact Wentzel--Kramers--Brillouin (WKB) analysis~\cite{Voros:1983xx,Delabaere:1999xx,Iwaki:2014vad}, have made it possible to formulate these states in a mathematically rigorous and physically meaningful way.
Exact WKB analysis has been successfully applied to a wide range of differential equations arising in physical systems:
thermodynamic Bethe ansatz equation~\cite{Ito:2018eon,Ito:2019jio,Emery:2020qqu,Ito:2023cyz,Ito:2024nlt},
Painl\'eve equation~\cite{Basar:2015xna,Iwaki:2019zeq,Hollands:2019wbr,Imaizumi:2020fxf,vanSpaendonck:2022kit,DelMonte:2022kxh},
supersymmetry (e.g. Seiberg--Witten curve)~\cite{Kashani-Poor:2015pca,Kashani-Poor:2016edc,Ashok:2016yxz,Ashok:2019gee,Yan:2020kkb,Imaizumi:2021cxf,Grassi:2021wpw,Bianchi:2021xpr,Alim:2022oll,Imaizumi:2022qbi,Imaizumi:2022dgj},
and cosmology~\cite{Enomoto:2020xlf,Enomoto:2021hfv,Enomoto:2022mti,Enomoto:2022nuj,Honda:2024aro}.
(See also Refs.~\cite{Taya:2020dco,Bucciotti:2023trp,Suzuki:2023slp}.)
Also, recently exact WKB analysis---serving as a \textit{nonperturbative formulation of QM}---has been applied to various systems, such as double-well potential~\cite{Sueishi:2020rug},
periodic-well potential~\cite{Sueishi:2021xti},
supersymmetric QM~\cite{Kamata:2021jrs},
and $PT$-symmetric QM~\cite{Kamata:2023opn,Kamata:2024tyb}\footnote{%
For more mathematical studies on $PT$-symmetric or other holomorphic functions, see Refs.~\cite{Nikolaev:2008,Nikolaev:2021xzt,Nikolaev:2024}.}.
These developments naturally raise the question: To what extent can such nonperturbative methods deepen our understanding of quasi-stable quantum systems?

Previous research has focused on an eventually stable system
while there exist many unstable systems of interest
that are quasi-stationary but not stable after decay.
A good example is a resonant state.
Resonance appears quite universally across various quantum many-body systems: scattering theory, nuclear physics, open quantum systems, and non-Hermitian QM.\footnote{For black hole physics, resurgence for resonance also occurs in quasi-normal modes. See e.g., Refs.~\cite{Miyachi:2025ptm,Hatsuda:2019eoj,Hatsuda:2021gtn,Matyjasek:2019eeu,Decanini:2011eh,Alfaro:2024tdr,Motl:2003cd}.}
Such a quasi-stable state is not normalizable.
To describe it, we often use analytic continuation or regularization such as the complex scaling method (CSM) and so on.
However, many aspects remain inexplicable because the probability becomes a complex number~\cite{Berggren:1996} and a transition cross-section to a resonant state is quite subtle~\cite{Berggren:1971,Berggren:1978}.
The CSM~\cite{Myo:2014ypa} seems to work well
but is formulated by sophisticated mathematical techniques~\cite{Aguilar:1971ve,Balslev:1971vb}.
Also, it is not transparent how to relate it to other methods such as Zel'dovich regularization~\cite{Zel'dovich:1961,Berggren:1968zz} and rigged Hilbert space~\cite{Rafael:2012,Antoniou:2001,Zhao:1995}.

In this paper, we investigate quasi-stationary quantum systems based on the exact WKB analysis as a nonperturbative framework.
In particular, we aim to understand resonance in this context.
As an example, we focus on the inverted Rosen--Morse potential~\cite{PhysRev.42.210,Zhao:1995,Landau:1991wop}.
This system gives rise to the so-called barrier resonance~\cite{Ryaboy:1993} but it is not intuitively clear why this happens.
We will give a simple picture from the analyticity.
This model is also exactly solvable;
we calculate the exact solution of resonant states\footnote{%
In supersymmetric QM, the system with the (noninverted) Rosen--Morse potential is a supersymmetric partner of the harmonic oscillator.
Then, this allows the derivation of exact solutions;
when the inverted case, the corresponding solution is given by the analytic continuation of this explicit expression.}
and compare the consistency of complex energies of resonance with predictions from the exact WKB analysis.
Notably, this provides a novel application of the method to a non-polynomial potential.

This paper is organized as follows:
We first review resonance itself in Sect.~\ref{sec:resonance}.
At first sight, this is given as pole singularities of an S-matrix.
Then some techniques will be introduced to use later.
Section~\ref{sec:ewkb} introduces the exact WKB analysis.
We also consider the cubic potential as a toy model;
to derive the quantization condition and understand some kind of decay are required.
In Sect.~\ref{sec:view_res}, we give a simple and general explanation of resonance (including barrier resonance) from the analyticity based on the exact WKB perspective.
In Sect.~\ref{sec:rm}, we consider the inverted Rosen--Morse potential from the traditional approach and the exact WKB method.
We will see the consistency of those complex energies.
Section~\ref{sec:conclusion} is devoted to the conclusion.

\section{Resonant states in scattering theory}\label{sec:resonance}
\begin{figure}[t]
 \centering
 \begin{subfigure}{0.45\columnwidth}
  \centering
  \begin{tikzpicture}[x=0.9mm,y=0.9mm,>=latex]
   \draw[->,line width=2pt] (-30,0) -- (30,0);
   \draw[->,line width=1pt] (0,-28) -- (0,28);
   \draw[line width=1pt] (-25,20) -- (-20,20) -- (-20,25);
   \draw[line width=1pt] (-22.5,22.5) node {$k$};
   \draw[line width=1pt,blue] (8,-5) circle (2);
   \draw[line width=1pt,blue] (16,-8) circle (2);
   \draw[line width=1pt,blue] (24,-13) circle (2);
   \fill (0,10) circle (2);
   \fill (0,18) circle (2);
   \draw[<-] (1.8,19.5) -- (10,22) node[right] {Bound};
   \draw[<-,blue] (16,-10.5) -- (15,-15) node[below] {Resonance};
  \end{tikzpicture}
  \caption{Complex $k$-plane}
  \label{fig:pole-a}
 \end{subfigure}
 \begin{subfigure}{0.45\columnwidth}
  \centering
  \begin{tikzpicture}[x=0.9mm,y=0.9mm,>=latex]
   \draw[line width=1pt] (22,25) .. controls (22,0) and (30,0) .. (30,-25);
   \draw[line width=1pt] (30,28) .. controls (30,0) and (22,0) .. (22,-28);
   \draw[line width=1pt] (22,25) -- (-30,25) -- (-30,-28) -- (22,-28);
   \draw[line width=1pt] (30,28) -- (-22,28) -- (-22,25);
   \draw[line width=1pt] (30,-25) -- (22,-25);
   \draw[line width=1pt] (-27,0) -- (-4,0);
   \draw[->,line width=2pt] (-4,0) -- (26,0);
   \draw[->,line width=1pt] (-4,-24) -- (-4,24);
   \draw[line width=1pt] (-26,17) -- (-21,17) -- (-21,22);
   \draw[line width=1pt] (-23.5,19.5) node {$E$};
   \draw[blue,dashed,line width=1pt] (2,-5) circle (2);
   \draw[blue,dashed,line width=1pt] (10,-10) circle (2);
   \draw[blue,dashed,line width=1pt] (18,-20) circle (2);
   \fill (-13,0) circle (2);
   \fill (-23,0) circle (2);
   \draw[<-] (-23,2) -- (-22.5,5) node[above] {Bound};
   \draw[blue,<-] (9.5,-12) -- (8,-20) node[below] {Resonance};
  \end{tikzpicture}
  \caption{Complex $E$-plane}
  \label{fig:pole-b}
 \end{subfigure} 
 \caption
 {Distribution of S-matrix poles in (a) the complex $k$-plane and (b) the complex $E$-plane. In the right panel~(b), the bound states exist in the first Riemann surface, while the resonant states appear in the second Riemann surface.}
 \label{fig:pole}
\end{figure}
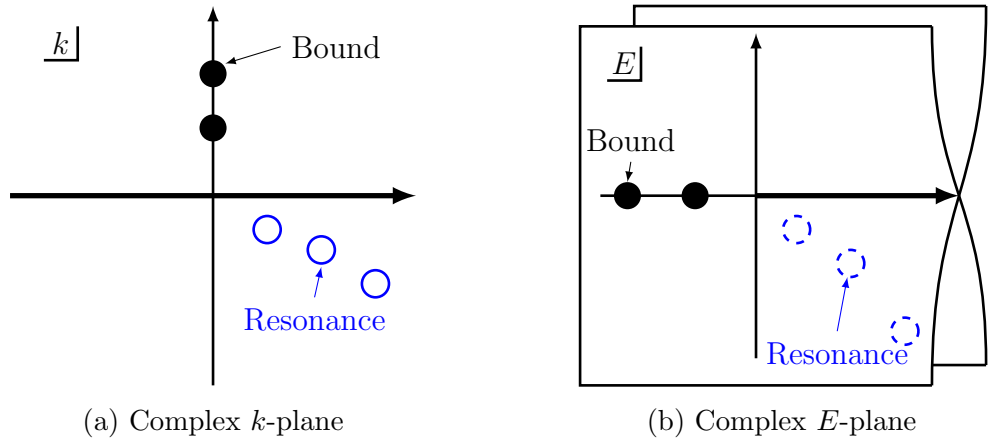

Figure~\ref{fig:pole-a} shows that bound and resonant states appear as poles of an S-matrix and are characterized by the complex wave number~\cite{Taylor:1972}. 
Resonances exist in the fourth quadrant of the complex $k$-plane.
Bound states are distributed along the negative region of the real axis in the complex $E$-plane, while resonant states are in the fourth quadrant as shown in Fig.~\ref{fig:pole-b}.
Note that bound and resonant states exist in the first and second Riemann surfaces, respectively, of the complex $E$-plane. A resonant state has complex energy, where the real part is the resonant energy and the imaginary part is the decay width. Resonance is produced by the scattering of multiple particles. If the decay width is narrow, a resonant state is observed as a sharp peak in the energy spectrum of the cross-section. 

The singular nature of a resonant state, which is the divergence of the wave function in the asymptotic region, causes difficulty in investigating the state. In the $3$D space, the radial wave function $\varphi(r)$ of a resonant state with $k=k_{r}-ik_{i}$ ($k_{r}$, $k_{i}\in\mathbb{R}^{+}$) has the following asymptotic form:
\begin{align}
 \varphi(r) \xrightarrow[r\to\infty]{}e^{ikr} = e^{ik_{r}}e^{k_{i}r},
\end{align}
which diverges due to the factor $e^{k_{i}r}$. This property also leads to the divergence of the norm, $||\varphi(r)||^2$. To obtain a finite value of the norm, the Zel'dovich regularization~\cite{Zel'dovich:1961} is introduced as
\begin{align}
 ||\varphi(r)||^2\equiv\lim_{\varepsilon\to+0}\int_{0}^{\infty} dr\, e^{-\varepsilon r^2}\varphi(r)^2r^2 .
\end{align}
Using this regularization, the orthogonality and completeness properties of the resonant states have been proved by Berggren~\cite{Berggren:1968zz}. It should be noted that the square of the wave function is integrated with the norm of the resonant state.

Recently, the CSM~\cite{Aguilar:1971ve,Balslev:1971vb,Myo:2014ypa} has been developed as another way to overcome the divergence of the norm and has been applied to describe resonant states of many-body systems, such as nuclei, atoms, and molecules. In the CSM, the coordinate $r$ is transformed as $r\to e^{i\theta}r$ with a parameter $\theta$. The ABC theorem argues that resonance, which is obtained from a Hamiltonian with this transformation, is square integrable. This can be naively understood as follows: In the CSM, the asymptotic form of the resonant wave function is transformed as
\begin{align}
 e^{ikr} \rightarrow e^{i(k_{r}-ik_{i})re^{i\theta}}=e^{(-k_{r}\sin\theta+k_{i}\cos\theta)r}e^{i(k_{r}\cos\theta+k_{i}\sin\theta)r},
\end{align}
which converges when $\theta>\tan^{-1}(k_{i}/k_{r})$.
The CSM thus enables a square-integrable treatment of resonant states by rotating the coordinate into the complex plane and has been successfully applied to many-body systems. Complementary approaches, such as the Gamov shell model~\cite{Michel:2009}, have also been developed to tackle the divergence problem inherent in resonant states.

\section{Exact WKB analysis and decay rate}\label{sec:ewkb}
\subsection{Brief introduction of exact WKB analysis}
The $1$D Schr\"odinger equation can be written as
\begin{align}
    \left[-\frac{\hbar^2}{2}\frac{d^2}{dx^2}+V(x)\right] \psi(x)
    = E \psi(x) .
\end{align}
For simplicity, we set
\begin{align}
    \left[-\frac{d^2}{dx^2}+\hbar^{-2}Q(x)\right] \psi(x) = 0,
    \qquad Q(x) = 2\left[V(x)-E\right] .
\end{align}
The WKB solution is an ansatz defined by a formal power series,
\begin{align}
    \psi(x,\hbar) &= e^{\int^x dx'\, S(x',\hbar)} ,&
    S(x,\hbar) &= \sum_{i=-1}^{\infty}\hbar^{i} S_{i}(x) .
\end{align}
Substituting $S(x,\hbar)$ into the Schr\"odinger equation,
we have the nonlinear Riccati equation;
we can obtain the recursive equation of $S_i$.
The leading order of the recursive equation is given by
\begin{align}
    S_{-1}(x) = S_{-1}^{\pm}(x) \equiv \pm \sqrt{Q(x)} .
\end{align}
It is known that (i) $S_i^{-}=(-1)^i S_i^{+}$,
(ii) $S^{\pm}=\pm S_{\mathrm{odd}} + S_{\mathrm{even}}$,
where $S_{\mathrm{odd}}=\sum_{i=-1,1,3,\dots}\hbar^i S_i^{+}$
and $S_{\mathrm{even}}=\sum_{i=0,2,4,\dots}\hbar^i S_i$,
(iii) $S_{\mathrm{even}}=-\frac{1}{2}\frac{d}{dx}\ln S_{\mathrm{odd}}$, and also (iv) the exact solution and approximation of the WKB wave function are
\begin{align}
    \psi(x,\hbar)^{\pm} = \frac{1}{\sqrt{S_{\mathrm{odd}}}}
    e^{\pm\int_{x_0}^x dx'\, S_{\mathrm{odd}}} 
    \sim \frac{1}{Q^{1/4}} e^{\pm\hbar^{-1}\int_{x_0}^x dx'\, \sqrt{Q}} + \dots.
\end{align}

The formal series is not necessarily convergent,
and one uses the so-called Borel resummation to give a finite value.
The essential point is the insertion of one,
\begin{align}
    1 = \frac{1}{\Gamma(n+\alpha)} \int_0^\infty dx\, e^{-x} x^{n+\alpha-1},
\end{align}
and exchange of the infinite sum and the above integral.
Now, the infinite sum becomes a convergent series,
while the integral may be singular for some parameters,
called Borel singularity.
Then, because the total value is ill-defined,
we pick up some analytically continued paths
that have an imaginary and nonperturbative nontrivial difference.
When the path jumps over the Borel singularity,
such a function suddenly changes (Stokes phenomenon).
In the exact WKB analysis, we can see such a condition
in terms of~$Q$ as
\begin{align}
    \im \hbar^{-1} \int^x dx' \sqrt{Q} = 0.
\end{align}
The curves determined by this condition are called Stokes curves,
and the structure is a Stokes graph.

\subsection{Application to unstable QM: Anharmonic oscillator with cubic potential}
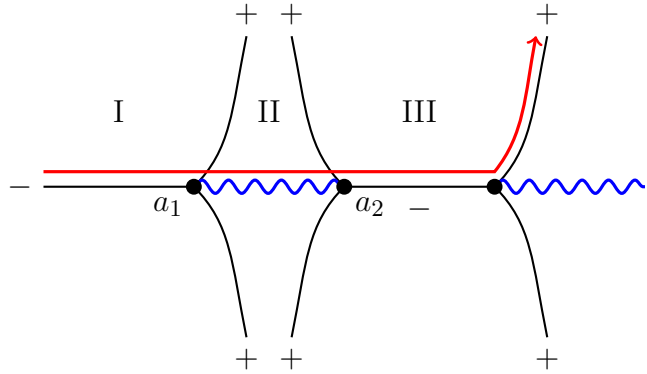
\begin{figure}
    \centering
\begin{tikzpicture}
    \draw[thick] (2,0) -- (0,0) node[left] {$-$};
    \draw[thick] (2,0) .. controls (2.5,0.5) and (2.5,1).. (2.7,2) node[above] {$+$};
    \draw[thick] (2,0) .. controls (2.5,-0.5) and (2.5,-1).. (2.7,-2) node[below] {$+$};
    \draw[thick] (4,0) .. controls (3.5,0.5) and (3.5,1).. (3.3,2) node[above] {$+$};
    \draw[thick] (4,0) .. controls (3.5,-0.5) and (3.5,-1).. (3.3,-2) node[below] {$+$};
    \draw[thick] (4,0) -- (6,0) node[midway,below] {$-$};
    \draw[thick] (6,0) .. controls (6.5,0.5) and (6.5,1).. (6.7,2) node[above] {$+$};
    \draw[thick] (6,0) .. controls (6.5,-0.5) and (6.5,-1).. (6.7,-2) node[below] {$+$};
    \draw[very thick,blue,snake it] (2,0) -- (4,0);
    \draw[very thick,blue,snake it] (6,0) -- (8,0);
    \fill (2,0) circle(3pt) node[below left] {$a_1$};
    \fill (4,0) circle(3pt) node [below right] {$a_2$};
    \fill (6,0) circle(3pt);
    \draw[very thick,red,->] (0,0.2) -- (6,0.2) .. controls (6.4,0.7) and (6.4,1.2).. (6.55,2);
    \node at (1,1) {I};
    \node at (3,1) {II};
    \node at (5,1) {III};
\end{tikzpicture}
    \caption{Stokes graphs for the cubic potential.
    The black solid curves are the Stokes curves
    and black points are the turning points connecting three Stokes curves.
    The blue wavy lines are branch cuts.
    The red arrowed curve is a possible physical path on which the wave function is defined.}
    \label{fig:cubic}
\end{figure}

As an example, let us consider the cubic potential, $V(x) = x^2(1-x)$.
This is not well-defined but is pedagogical for our purpose.
The Stokes graph for~$0<E<\max V(x>0)$ is shown in Fig.~\ref{fig:cubic}.
The black solid curves are the Stokes curves.
The index of each curve $\pm$ implies that $\psi^{\pm}$ increases exponentially and
\begin{align}
    \re \hbar^{-1} \int^x dx' \sqrt{Q} \gtrless 0.
\end{align}
Three Stokes curves gather at one point, called the turning point, which is depicted as a black point.
The blue wavy lines are branch cuts.

Due to the unbounded potential, it is difficult to define physical phenomena only on~$x\in(-\infty,\infty)$.
For example, let us choose the red arrowed path and then see what happens.
At first, near $x=-\infty$, $\psi^{-}$ sufficiently decreases and so is normalizable.
When we jump over the Stokes curve with the index~$+$ starting from the same turning point (I$\to$II), the Stokes phenomenon gives rise to
\begin{align}
    \begin{pmatrix}
      \psi^{+}_{\mathrm{I}} \\ \psi^{-}_{\mathrm{I}}
    \end{pmatrix}
    =
    M
    \begin{pmatrix}
      \psi^{+}_{\mathrm{II}} \\ \psi^{-}_{\mathrm{II}}
    \end{pmatrix} ,
    \label{eq:connection}
\end{align}
where $M$ depends on the rotation of the path around the turning point,
\begin{align}
    M =
    \begin{cases}
        M_{+} & \text{if anticlockwise for the index~$+$,} \\
        M_{+}^{-1}& \text{if clockwise for the index~$+$,} \\
        M_{-} & \text{if anticlockwise for the index~$-$,} \\
        M_{-}^{-1} &\text{if clockwise for the index~$-$,}
    \end{cases}
\end{align}
and
\begin{align}
    M_{+} &=
    \begin{pmatrix}
        1 & i \\ 0 & 1
    \end{pmatrix}, &
    M_{-} &=
    \begin{pmatrix}
        1 & 0 \\ i & 1
    \end{pmatrix} .
\end{align}
The above matrix is called the monodromy matrix.
Inside region II, we need to connect the different turning points, $a_1$ and $a_2$.
The wave function is related to each other by the factor,
\begin{align}
    e^{\pm\int_{a_1}^{a_2}dx\, S_{\mathrm{odd}}}
\end{align}
Finally, in the exchange of regions II and III, we again use the connection rule~\eqref{eq:connection}.

If this system is just the harmonic oscillator, $V(x)=x^2$,
the above strategy completes our task.
For the wave function to converge near~$x=\infty$,
we have the quantization condition
\begin{align}
    1 + A = 0, \qquad
    A = e^{\oint dx\, S_{\mathrm{odd}}} .
\end{align}
Here the cycle, $A$, comes from the connection between the turning points~$a_1$ and $a_2$.
Actually, by analytically continuing,
$A$ can be rewritten as the integral around the point of infinity.
That is, $A=e^{2\pi i \hbar^{-1} E}$;
we have the usual quantized energy spectrum~$E=\hbar\left(n+\frac{1}{2}\right)$ with $n\in\mathbb{Z}_{\geq0}$.

Now, for the cubic potential, the analytical continuation should be stuck because of the branch cut attached to the point of infinity.
However, a path on the upper-half plane must differ from that on the lower-half plane.
This reproduces the usual result of some kind of tunneling analysis.
This section is now finished but we will see a concrete result
being identical to the exact solution.

\section{Resonant states in exact WKB analysis}\label{sec:res_ewkb}
\subsection{What is a resonant state from the viewpoint of exact WKB?}\label{sec:view_res}
\begin{figure}
    \centering
    \begin{subfigure}{0.48\columnwidth}
        \centering
        \includegraphics[width=\columnwidth]{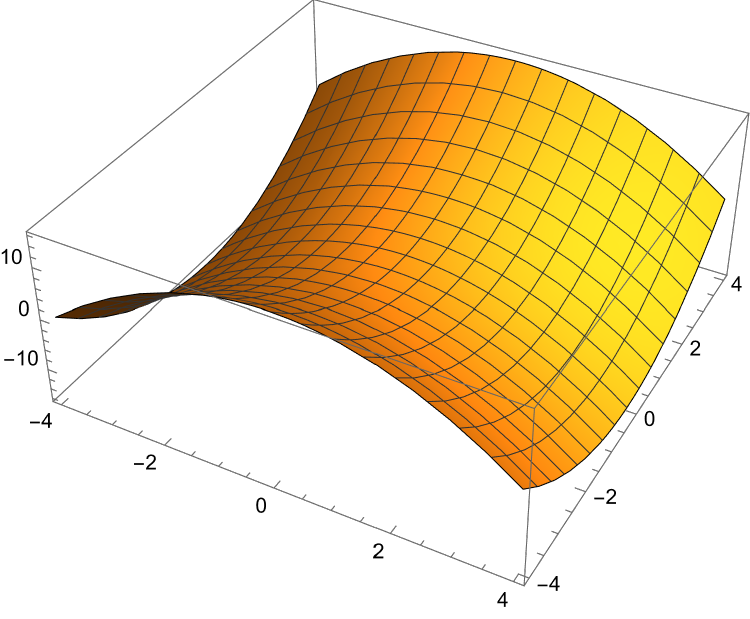}
        \caption{Real part}
    \end{subfigure}\hspace*{1em}
    \begin{subfigure}{0.48\columnwidth}
        \includegraphics[width=\columnwidth]{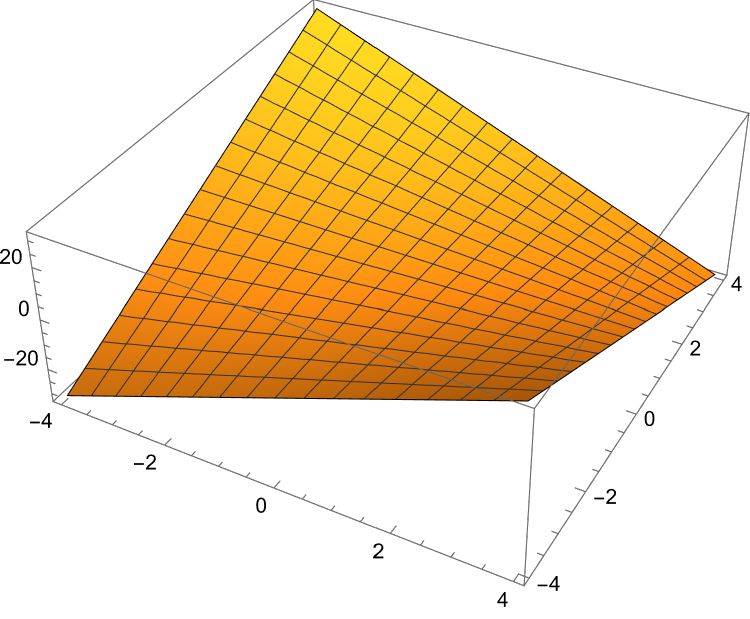}
        \caption{Imaginary part}
    \end{subfigure}
    \caption{Potential structure of harmonic oscillator, $-z^2$, in complex plane}
    \label{fig:V_harmonic}
\end{figure}

\begin{figure}
    \centering
    \begin{subfigure}{0.48\columnwidth}
        \centering
        \includegraphics[width=\columnwidth]{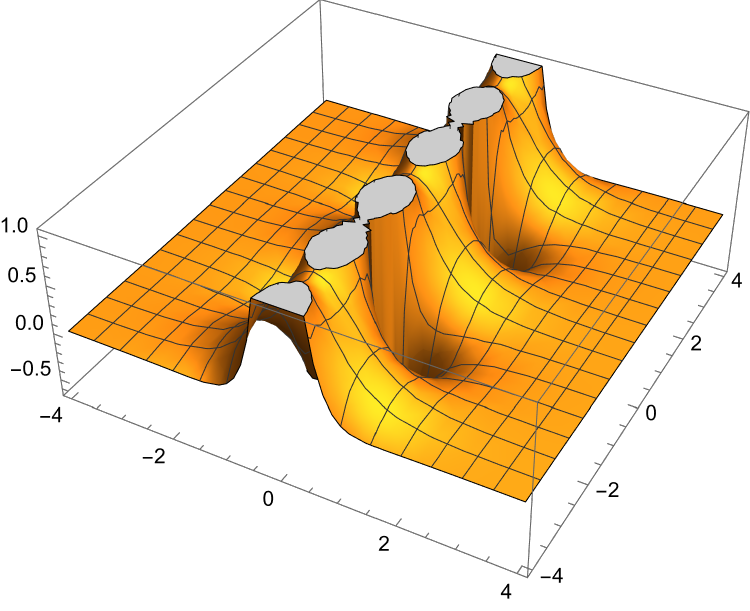}
        \caption{Real part}
    \end{subfigure}\hspace*{1em}
    \begin{subfigure}{0.48\columnwidth}
        \includegraphics[width=\columnwidth]{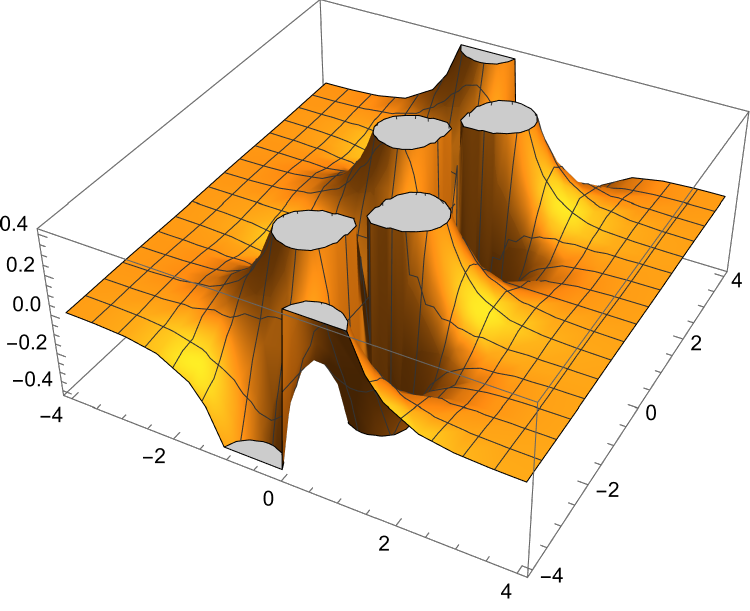}
        \caption{Imaginary part}
    \end{subfigure}
    \caption{Potential structure of $1/\cosh^2{z}$ in complex plane}
    \label{fig:V_cosh}
\end{figure}

Given a potential $V(x)$ as a function of~$x\in\mathbb{R}$,
let us complexify the coordinate as $z\in\mathbb{C}$.
In the complex plane~$z$,
if the potential $\re V(z)$ possesses local minima, all of which are not stable physical states,
then the Hilbert space of the system includes resonant states.
This is because observations from the exact WKB analysis, i.e. structures of Stokes curves, are sensitive to the change of signs of~$V(z)-E$.
For instance, in an unusual harmonic oscillator, $V(x)=-x^2$,
one finds that in~Fig.~\ref{fig:V_harmonic} there is no local minimum; no resonant state exists.
On the other hand, some kind of potential has the local minimum which is not a true vacuum, such as in~$1/\cosh^2(x-a)+1/\cosh^2(x+a)$; there are resonant states.
Additionally, it is known that for the potential,
$V(x)=1/\cosh^2{x}$,
we observe nontrivial complex energies of resonance.
This is the so-called barrier resonance,
which traps the wave function into the bump.
In fact, in~Fig.~\ref{fig:V_cosh}, we can see the existence of local minima in $\re V(z)$ with~$\im z\neq0$.

In what follows, we focus on an exactly solvable model, where we can see barrier resonant states explicitly.
Also, from the above picture, we will have the same complex energies by using the exact WKB analysis.

\subsection{Inverted Rosen--Morse potential}\label{sec:rm}
\subsubsection{Exact solution}
Let us consider the inverted Rosen--Morse potential defined by
\begin{align}
    \left[-\frac{\hbar^2}{2m} \frac{d^2}{dx^2} + \frac{U_{0}}{\cosh^2 \beta x}\right] \psi(x)
    = E \psi(x) ,
    \qquad U_{0}>0.
\end{align}
This Schr\"odinger equation is exactly solvable as follows:
By using $\xi = \tanh \beta x$ and
\begin{align}
    k = \frac{\sqrt{2mE}}{\hbar}, \quad
    -\frac{2mU_{0}}{\beta^2\hbar^2} = s(s+1), \quad
    s=\frac{1}{2}\left( -1+\sqrt{1-\frac{8mU_{0}}{\beta^2 \hbar^2}} \right)
\end{align}
the equation can be rewritten as
\begin{align}
    \frac{d}{d\xi}\left[ (1-\xi^2) \frac{d\psi}{d\xi} \right] +
    \left[ s(s+1) - \left(\frac{ik}{\beta}\right)^{2} \frac{1}{1-\xi^2} \right]\psi = 0 .
\end{align}
This is a general Legendre equation.
After changing variables as
\begin{align}
    \psi(x) = (1-\xi^2)^{\frac{ik}{2\beta}} w(\xi)
\end{align} 
and $\frac{1}{2}(1-\xi)=u$, one finds the hypergeometric differential equation,
\begin{align}
    u(1-u)\frac{d^2}{du^2}w + \left( \frac{ik}{\beta} + 1 \right)(1-2u)\frac{d}{du}w
    - \left( \frac{ik}{\beta} - s \right) \left( \frac{ik}{\beta} + s + 1 \right)w = 0 .
\end{align}

It is known that there are two independent solutions to the above equation.
One solution is finite at $\xi=1$, i.e. $x=\infty$, while another one is divergent there.
The former is given by
\begin{align}
    \psi(x) = 
    (1-\xi^2)^{-\frac{ik}{2\beta}}
    F\left( -\frac{ik}{\beta}-s, -\frac{ik}{\beta}+s+1, -\frac{ik}{\beta}+1, \frac{1-\xi}{2} \right) ,
 \label{eq:rm_sol}
\end{align}
where $F$ is the Gaussian hypergeometric function.

\subsubsection{Traditional approach to resonance}\label{sec:rm_resonance}
The exact solution~\eqref{eq:rm_sol} has complex energies.
To see this from the traditional viewpoint,
the asymptotic behavior of the solution is of importance.
At first, in the case that $x\to\infty$, we see
\begin{align}
    (1-\xi^2)^{-\frac{ik}{2\beta}}
    &\to 
    4^{-\frac{ik}{2\beta}} e^{ikx}, 
\end{align}
and
\begin{align}
    F\left( -\frac{ik}{\beta}-s, -\frac{ik}{\beta}+s+1, -\frac{ik}{\beta}+1, \frac{1-\xi}{2} \right)  
    \to 1 .
\end{align}
Thus, we have the asymptotic form of the exact solution at~$x\to\infty$
\begin{align}
    \psi(x) \rightarrow  4^{-\frac{ik}{2\beta}} e^{ikx} .
\end{align}
Here, the above expression means the transmitted wave.
Next, if $x\to-\infty$, the overall factor becomes
\begin{align}
    (1-\xi^2)^{-\frac{ik}{2\beta}}
    &\to
    (-4)^{-\frac{ik}{2\beta}} e^{-ikx} .
\end{align}
On the other hand, noting that
\begin{align}
    &F\left( -\frac{ik}{\beta}-s, -\frac{ik}{\beta}+s+1, -\frac{ik}{\beta}+1, \frac{1-\xi}{2} \right) 
    \notag \\
    &=
    \frac{\Gamma\left(1-\frac{ik}{\beta}\right) \Gamma\left(\frac{ik}{\beta}\right)}{\Gamma(1+s) \Gamma(-s)}
    F\left( -\frac{ik}{\beta}-s, -\frac{ik}{\beta}+s+1, -\frac{ik}{\beta}+1, \frac{1+\xi}{2} \right) 
    \notag \\
    &\qquad +
    \frac{\Gamma\left(1-\frac{ik}{\beta}\right) \Gamma\left(-\frac{ik}{\beta}\right)}
    {\Gamma\left(-\frac{ik}{\beta}-s\right) \Gamma\left(-\frac{ik}{\beta}+s+1\right)}
    \left(\frac{1+\xi}{2}\right)^{\frac{ik}{a}}
    F\left( 1+s, -s, 1+\frac{ik}{\beta}, \frac{1+\xi}{2} \right) 
\end{align}
the hypergeometric function behaves as
\begin{align}
    &F\left( -\frac{ik}{\beta}-s, -\frac{ik}{\beta}+s+1, -\frac{ik}{\beta}+1, \frac{1-\xi}{2} \right) \notag\\
    &\to
    \frac{\Gamma\left(1-\frac{ik}{\beta}\right) \Gamma\left(\frac{ik}{\beta}\right)}{\Gamma(1+s) \Gamma(-s)}
    +
    \frac{\Gamma\left(1-\frac{ik}{\beta}\right) \Gamma\left(-\frac{ik}{\beta}\right)}
    {\Gamma\left(-\frac{ik}{\beta}-s\right) \Gamma\left(-\frac{ik}{\beta}+s+1\right)}
    e^{2ikx} .
\end{align}
Therefore, we have in the $x\to-\infty$ region
\begin{align}
    \psi(x) &\to
    (-4)^{-\frac{ik}{2\beta}} e^{-ikx} 
    \frac{\Gamma\left(1-\frac{ik}{\beta}\right) \Gamma\left(\frac{ik}{\beta}\right)}{\Gamma(1+s) \Gamma(-s)} \notag\\&\qquad
    +
    (-4)^{-\frac{ik}{2\beta}} e^{ikx} 
    \frac{\Gamma\left(1-\frac{ik}{\beta}\right) \Gamma\left(-\frac{ik}{\beta}\right)}
    {\Gamma\left(-\frac{ik}{\beta}-s\right) \Gamma\left(-\frac{ik}{\beta}+s+1\right)},
\end{align}
where the first term on the right-hand side indicates the reflected wave, and the second term is the incident wave.
Imposing purely outgoing boundary conditions (Siegert boundary condition~\cite{Siegert:1939}) yields the resonance condition, where the coefficient of the incident wave~$e^{ikx}$ vanishes.
Hence, $\Gamma\left(-\frac{ik}{\beta}-s\right)$ or $\Gamma\left(-\frac{ik}{\beta}+s+1\right)$ should be divergent, and hence those Gamma functions become $\Gamma(-n)$ with a nonnegative integer $n$ (i.e. $n=0$, $1$, $2$, \dots). Therefore, this leads to discrete complex energies as follows:
\begin{itemize}
 \item The case that $\Gamma\left(-\frac{ik}{\beta}+s+1\right)$ diverges.\\
  The divergence point of the Gamma function is expressed as
  \begin{align}
    -\frac{ik}{\beta}+s+1=-n, \qquad \text{$n=0$, $1$, $2$, \dots} .
  \end{align}
  Solving this equation for $k$, we obtain
  \begin{align}
    k=\frac{\beta}{2}
    \left[-i\sqrt{\frac{8mU_{0}}{\beta^2 \hbar^2}-1}-i(2n+1)\right] .
  \end{align}
  Here, if the inequality, $1<8mU_{0}/(\beta^2 \hbar^2)$, is satisfied, the complex wave number and the complex energy are given by 
  \begin{align}
  \label{eq:wn-Res}
    k^{\rm R}_{n}=\frac{\beta}{2}
    \left[\sqrt{\frac{8mU_{0}}{\beta^2 \hbar^2}-1}-i(2n+1)\right] 
  \end{align}
  and
  \begin{align}
    E^{\rm R}_{n}=\frac{\hbar^2 \beta^2}{8m}
    \left[\sqrt{\frac{8mU_{0}}{\beta^2 \hbar^2}-1}-i(2n+1)\right]^2,
  \end{align}
  respectively. The wave number is in the fourth quadrant of the complex $k$-plane. Therefore, this solution corresponds to a resonant state.  \\
 \item The case that $\Gamma\left(-\frac{ik}{\beta}-s\right)$ diverges.\\
  In the same way as the above case, we can obtain the divergence point of the Gamma function as
  \begin{align}
    -\frac{ik}{\beta}-s=-n, \qquad \text{$n=0$, $1$, $2$, \dots},
  \end{align}
  which gives 
  \begin{align}
   \label{eq:wn-ARes}
    k^{\rm AR}_{n}=\frac{\beta}{2}
    \left[-\sqrt{\frac{8mU_{0}}{\beta^2 \hbar^2}-1}-i(2n+1)\right]
  \end{align}
  and 
  \begin{align}
    E^{\rm AR}_{n}=\frac{\hbar^2 \beta^2}{8m}
    \left[\sqrt{\frac{8mU_{0}}{\beta^2 \hbar^2}-1}+i(2n+1)\right]^2 .
  \end{align}
  The wave number is in the third quadrant of the complex $k$-plane. Equations \eqref{eq:wn-Res} and \eqref{eq:wn-ARes} satisfy the following relation:
  \begin{align}
    k^{\rm R}_{n} = - (k^{\rm AR}_{n})^{*},
  \end{align}
  which means that the state with $k^{\rm AR}_{n}$ is a time-reversal state of the resonant state, the so-called ``antiresonant state.''
\end{itemize}

\subsubsection{Stokes graph}\label{sec:rm_stokes}
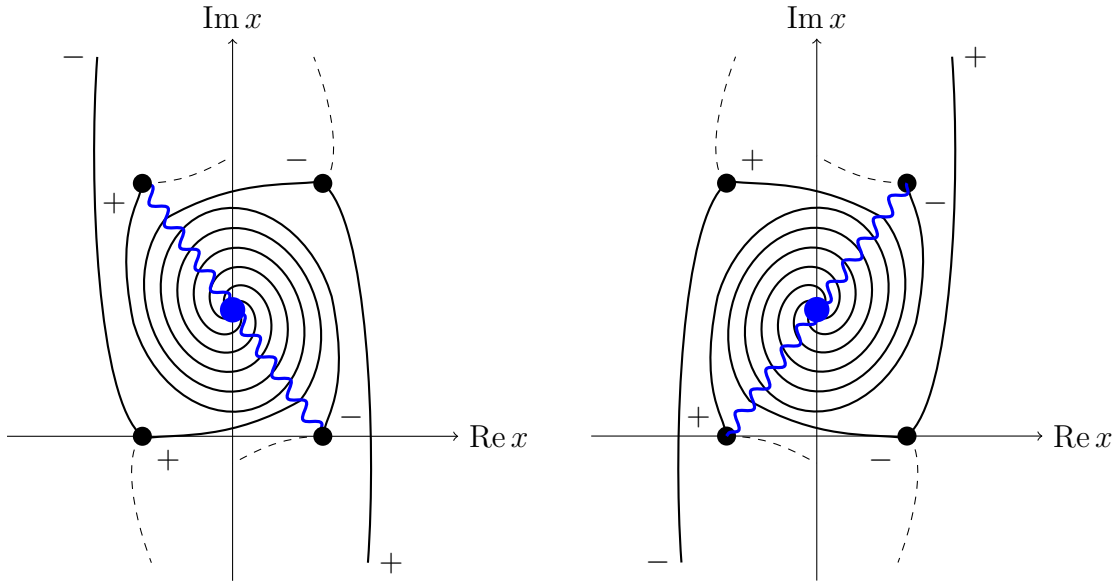
\begin{figure}[t]
\centering
\begin{tikzpicture}[scale=1.2]
  \draw[->] (-2.5,-1.4) -- (2.5,-1.4) node[right] {$\re x$};
  \draw[->] (0,-3) -- (0,3) node[above] {$\im x$};
  \draw[thick,domain=0:520,samples=100,variable=\t,rotate=20] plot ({-0.002*\t*sin(\t)}, {-0.0018*\t^1.05*cos(\t)}) ..controls (0.5,1.35) and (1,1.1).. (1.4,1) node[above left] {$-$};
  \draw[thick,domain=0:-520,samples=100,variable=\t,rotate=20] plot ({0.002*\t*sin(\t)}, {-0.0018*\t^1.05*cos(\t)}) ..controls (-0.5,-1.35) and (-1,-1.1).. (-1.4,-1) node[below right] {$+$};
  \draw[thick,domain=0:550,samples=100,variable=\t,rotate=110] plot ({-0.0018*\t^1.05*sin(\t)}, {-0.002*\t*cos(\t)}) ..controls (1.5,0.9) and (1.5,0.4).. (1.7,0.5) node[below left] {$+$};
  \draw[thick,domain=0:-550,samples=100,variable=\t,rotate=110] plot ({0.0018*\t^1.05*sin(\t)}, {-0.002*\t*cos(\t)}) ..controls (-1.5,-0.9) and (-1.5,-0.4).. (-1.7,-0.5) node[above right] {$-$};
  \draw[dashed] (1,1.4) ..controls (1.15,1.5) and (1.2,2) .. (0.9,2.8);
  \draw[dashed] (-1,-1.4) ..controls (-1.15,-1.5) and (-1.2,-2) .. (-0.9,-2.8);
  \draw[dashed] (-1,1.4)   ..controls (-0.5,1.4)   and (-0.2,1.6)  .. (0,1.7);
  \draw[dashed] (1,-1.4) ..controls (0.5,-1.4) and (0.2,-1.6) .. (0,-1.7);
  \draw[thick] (1,1.4) ..controls (1.5,1) and (1.6,-1.4) .. (1.5,-2.8) node[right] {$+$};
  \draw[thick] (-1,-1.4) ..controls (-1.5,-1) and (-1.6,1.4) .. (-1.5,2.8) node[left] {$-$};
  \fill[blue] (0,0) circle(4pt);
  \draw[very thick,blue, snake it] (-1,1.4) -- (1,-1.4);
  \fill (-1,1.4) circle(3pt); \fill (1,1.4) circle(3pt);
  \fill (1,-1.4) circle(3pt); \fill (-1,-1.4) circle(3pt);
\end{tikzpicture}
\hspace{1em}
\begin{tikzpicture}[scale=1.2]
  \draw[->] (-2.5,-1.4) -- (2.5,-1.4) node[right] {$\re x$};
  \draw[->] (0,-3) -- (0,3) node[above] {$\im x$};
  \fill (-1,1.4) circle(3pt); \fill (1,1.4) circle(3pt);
  \fill (1,-1.4) circle(3pt); \fill (-1,-1.4) circle(3pt);
  \draw[thick,domain=0:-520,samples=100,variable=\t,rotate=-20] plot ({0.002*\t*sin(\t)}, {0.0018*\t^1.05*cos(\t)}) ..controls (-0.5,1.35) and (-1,1.1).. (-1.4,1) node[above right] {$+$};
  \draw[thick,domain=0:520,samples=100,variable=\t,rotate=-20] plot ({-0.002*\t*sin(\t)}, {0.0018*\t^1.05*cos(\t)}) ..controls (0.5,-1.35) and (1,-1.1).. (1.4,-1) node[below left] {$-$};
  \draw[thick,domain=0:-550,samples=100,variable=\t,rotate=-110] plot ({0.0018*\t^1.05*sin(\t)}, {0.002*\t*cos(\t)}) ..controls (-1.5,0.9) and (-1.5,0.4).. (-1.7,0.5) node[below right] {$-$};
  \draw[thick,domain=0:550,samples=100,variable=\t,rotate=-110] plot ({-0.0018*\t^1.05*sin(\t)}, {0.002*\t*cos(\t)}) ..controls (1.5,-0.9) and (1.5,-0.4).. (1.7,-0.5) node[above left] {$+$};
  \draw[dashed] (-1,1.4) ..controls (-1.15,1.5) and (-1.2,2) .. (-0.9,2.8);
  \draw[dashed] (1,-1.4) ..controls (1.15,-1.5) and (1.2,-2) .. (0.9,-2.8);
  \draw[dashed] (1,1.4)   ..controls (0.5,1.4)   and (0.2,1.6)   .. (0,1.7);
  \draw[dashed] (-1,-1.4) ..controls (-0.5,-1.4) and (-0.2,-1.6) .. (0,-1.7);
  \draw[thick] (-1,1.4) ..controls (-1.5,1) and (-1.6,-1.4) .. (-1.5,-2.8) node[left] {$-$};
  \draw[thick] (1,-1.4) ..controls (1.5,-1) and (1.6,1.4) .. (1.5,2.8) node[right] {$+$};
  \fill[blue] (0,0) circle(4pt);
  \draw[very thick,blue, snake it] (1,1.4) -- (-1,-1.4);
\end{tikzpicture}
\caption{Schematically illustrated Stokes graph for the inverted Rosen--Morse potential, $V(x)=1/\cosh^2x$.
\textbf{(Left)} $\im\hbar>0$, \textbf{(right)} $\im\hbar<0$.
The black points denote the turning points and the solid curves are the corresponding Stokes curves.
The dashed ones, also Stokes curves, mean the periodicity of~$\im x\in[-\pi/2,\pi/2]$ due to the cosine function on~$\re x=0$.
The blue point is the double pole and the blue wavy lines are branch cuts.}
\label{fig:stokes_graph}
\end{figure}

So far, we have derived the exact resonant solution and its energy.
Here, we proceed with the perspective of the exact WKB analysis.
Figure~\ref{fig:stokes_graph} shows the Stokes graph for the inverted Rosen--Morse potential, $V(x)=1/\cosh^2x$, with $0<E<1$.
The left and right panels are devoted to~$\im\hbar>0$ and $\im\hbar<0$, respectively.
The black points denote the turning points and the solid/dashed curves are the corresponding Stokes curves.
Note that the potential possesses the periodicity of~$\im x\in[-\pi/2,\pi/2]$ due to the cosine function on~$\re x=0$.
The blue point is the double pole and the blue wavy lines are branch cuts.

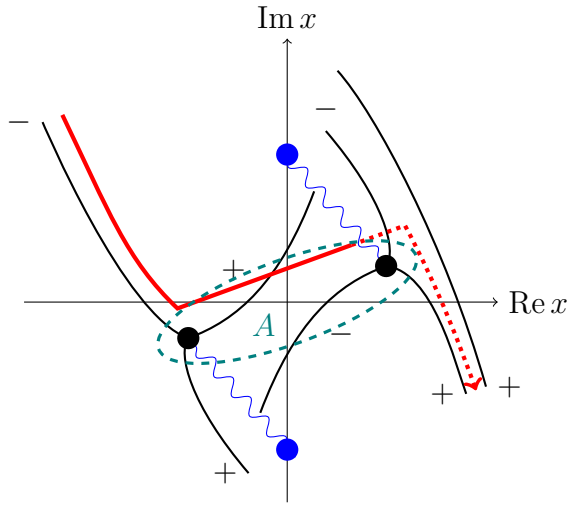
\begin{figure}[t]
\centering
\begin{tikzpicture}[scale=1.4]
  \draw[->] (-2.5,-1.4) -- (2,-1.4) node[right] {$\re x$};
  \draw[->] (0,-3.3) -- (0,1.1) node[above] {$\im x$};
  \draw[thick,rotate around={20:(0,-1.4)}] (1,-1.4) ..controls (1.15,-1.3) and (1.2,-0.8) .. (0.9,0) node[above] {$-$};
  \draw[thick,rotate around={20:(0,-1.4)}] (-1,-1.4) ..controls (-1.15,-1.5) and (-1.2,-2) .. (-0.9,-2.8) node[left] {$+$};
  \draw[thick,rotate around={20:(0,-1.4)}] (-1,-1.4)   ..controls (-0.5,-1.4)   and (0,-1.2)  .. (0.6,-0.5) node[midway,above left] {$+$};
  \draw[thick,rotate around={20:(0,-1.4)}] (1,-1.4) ..controls (0.5,-1.4) and (0,-1.6) .. (-0.6,-2.3) node[midway,below right] {$-$};
  \draw[thick,rotate around={20:(0,-1.4)}] (1.2,0.5) ..controls (1.4,0) and (1.6,-1.4) .. (1.5,-2.8) node[right] {$+$};
  \draw[thick,rotate around={20:(0,-1.4)}] (-1,-1.4) ..controls (-1.5,-1) and (-1.6,1) .. (-1.6,1) node[left] {$-$};
  \draw[thick,rotate around={20:(0,-1.4)}] (1,-1.4) ..controls (1.4,-1.6) and (1.3,-2.5) .. (1.3,-2.8) node[left] {$+$};
  \draw[ultra thick,red,rotate around={20:(0,-1.4)}] (-1.4,1) ..controls (-1.3,0) and (-1.3,-0.5) .. (-1,-1.1) -- (0.76,-1.1);
  \draw[ultra thick,dotted,red,->,rotate around={20:(0,-1.4)}] (0.76,-1.1) -- (1.3,-1.1) .. controls (1.4,-1.8) and (1.4,-2.4) .. (1.4,-2.8);
  \fill[blue] (0,0) circle(3pt); \fill[blue] (0,-2.8) circle(3pt);
  \draw[blue, snake it,rotate around={20:(0,-1.4)}] (0.43,0.01) -- (1,-1.4);
  \draw[blue, snake it,rotate around={20:(0,-1.4)}] (-0.43,-2.81) -- (-1,-1.4);
  \fill[rotate around={20:(0,-1.4)}] (1,-1.4) circle(3pt); \fill[rotate around={20:(0,-1.4)}] (-1,-1.4) circle(3pt);
  \draw[very thick,dashed,teal,rotate around={20:(0,-1.4)}] (0,-1.4) circle [x radius=1.3,y radius=0.4] node[below left] {$A$};
\end{tikzpicture}
\caption{Quantization condition with~$\im\hbar>0$ and~$\im E<0$.
The red path is an option to carry out the analytical continuation,
along which we can normalize the exact WKB solution.
The nontrivial cycle, $A$-cycle, is shown as the green dashed loop.}
\label{fig:stokes_graph_path}
\end{figure}

To see the resonant states, we choose a path of the analytic continuation from~$\im x\to\pm\infty$ to~$\im x\to\mp\infty$ along the Stokes curves depicted in~Fig.~\ref{fig:stokes_graph}.
Especially, because of normalizability,
we need to take $\psi_-$ for both regions of~$\im x\to\pm\infty$.
Such a path can be chosen as shown in Fig.~\ref{fig:stokes_graph_path}.
The red path in the figure implies the path in the above sense.
Note that the red solid curve is on the Riemann sheet depicted
while the red dashed curve is on another Riemann sheet after passing through the branch cut.

We make a remark on analogy to the CSM. From the ABC theorem, resonant states after complex scaling are some kind of bound states. It indicates that resonant states can be normalized in complexified asymptotic regions. In fact, in the Stokes graph in Fig.~\ref{fig:stokes_graph_path}, we observe that there is a path in the complex plane along which the wave function decays; we regard it as the normalizable path.

Figure~\ref{fig:stokes_graph_path} gives rise to the following quantization condition
along the red arrowed path:
\begin{align}
    1 - A = 0 .
    \label{eq:quantization_cond}
\end{align}
Here, $A$ is the nontrivial cycle, the $A$-cycle, shown as the green dashed loop.
From now on, let us calculate $A$
in the leading order of the exact WKB
and a completely exact analysis.

\subsubsection{Leading order}\label{sec:rm_leading}
In the leading order,
we carry out an integral of the WKB approximation between the turning points:
\begin{align}
    \int_{-\cosh^{-1}\sqrt{\frac{1}{E}}}^{\cosh^{-1}\sqrt{\frac{1}{E}}} dx \sqrt{2(V(x) - E)}
    = \sqrt{2} \int dx\sqrt{\frac{1}{\cosh^2 x} - E} .
\end{align}
Here we set almost all parameters as units for simplicity.
After some calculations, we find that
\begin{align}
    &\int_{-\cosh^{-1}\sqrt{\frac{1}{E}}}^{\cosh^{-1}\sqrt{\frac{1}{E}}} dx \sqrt{2(V(x) - E)}\notag\\
    &= - 2 \cosh (x) 
   \sqrt{\frac{-E-\tanh ^2(x)+1}{E \cosh (2 x)+E-2}}
   \notag\\&\qquad\times
   \left[
   \tanh^{-1} \left(\frac{\sqrt{2} \sinh (x)}{\sqrt{E \cosh (2 x)+E-2}}\right)
   -\sqrt{E} \sinh ^{-1}\left(\frac{\sinh(x)}{\sqrt{1-\frac{1}{E}}}\right)
   \right]
   \\
    &= \mp \sqrt{2} i
   \left[
   \tanh^{-1} \left(\frac{\sqrt{2} \sinh (x)}{\sqrt{E \cosh (2 x)+E-2}}\right)
   -\sqrt{E} \sinh ^{-1}\left(\frac{\sinh(x)}{\sqrt{1-\frac{1}{E}}}\right)
   \right]
   \\
   &= \pm \sqrt{2} \pi (1 + \sqrt{E}) .
\end{align}
Then, from the quantization condition~\eqref{eq:quantization_cond},
\begin{align}
    2 \sqrt{2} \pi (1 + \sqrt{E}) = 2\pi i n,
\end{align}
with $n\in\mathbb{Z}$.
Therefore, the leading complex energy is given by
\begin{align}
    E = \left( 1 - \frac{i n}{\sqrt{2}} \right)^2 .
\end{align}
This is quite suggestive because we observe resonance in the fourth quadrant even in the leading order of the exact WKB method.\footnote{For the usual (noninverted) Rosen--Morse potential, higher-order contribution from~$S_i$ vanishes. (This is the reason why it possesses shape invariance in the same way as the harmonic oscillator.) Then, one finds an exact solution. On the other hand, higher-order terms $S_i$ do not vanish for the inverted Rosen--Morse potential. We need some careful treatment.}

\subsubsection{Exact WKB analysis for resonance}\label{rm_ewkb}
In the above Stokes graphs, the exact solution on the normalizable path
is explicitly written by the exact solution~\eqref{eq:rm_sol}.
Note that another solution has opposite signs for Stokes curves.
This means that the WKB solution can be exactly given by
\begin{align}
    e^{\int_{x_0}^{x} dx' S_{\mathrm{odd}}}
    = \left.(1-\xi^2)^{-\frac{ik}{2\beta}}
    F\left( -\frac{ik}{\beta}-s, -\frac{ik}{\beta}+s+1, -\frac{ik}{\beta}+1, \frac{1-\xi}{2} \right)\right|_{\text{at $x$}},
\end{align}
where $x_0$ is a reference point.
The $A$-cycle, $e^{\oint S_{\mathrm{odd}}}$, between turning points
$\mp\cosh^{-1}\sqrt{U_0/E}$ can be realized by the exchange of reference points.
That is,
\begin{align}
    e^{2\int_{x_0}^{\cosh^{-1}\sqrt{U_0/E}} dx\, S_{\mathrm{odd}}}
    \times e^{-2\int_{x_0}^{-\cosh^{-1}\sqrt{U_0/E}} dx\, S_{\mathrm{odd}}}
    = e^{2\int_{-\cosh^{-1}\sqrt{U_0/E}}^{\cosh^{-1}\sqrt{U_0/E}} dx\, S_{\mathrm{odd}}} .
\end{align}
From the fact that $A=1$ for the existence of a normalizable solution, we see
\begin{align}
    \left[
    \frac{F\left( -\frac{ik}{\beta}-s, -\frac{ik}{\beta}+s+1, -\frac{ik}{\beta}+1, \frac{1-|\xi|}{2} \right)}
    {F\left( -\frac{ik}{\beta}-s, -\frac{ik}{\beta}+s+1, -\frac{ik}{\beta}+1, \frac{1+|\xi|}{2} \right)}
    \right]^2
    = 1 ,
    \label{eq:quant_cond_F}
\end{align}
where $|\xi|=\tanh(\beta\cosh^{-1}\sqrt{U_0/E})$.

Here, we use the formula of the hypergeometric function as
\begin{align}
    &F\left( -\frac{ik}{\beta}-s, -\frac{ik}{\beta}+s+1, -\frac{ik}{\beta}+1, \frac{1\pm|\xi|}{2} \right)\notag\\
    &= \mathfrak{A} F\left(-\frac{ik}{2\beta}-\frac{s}{2},-\frac{ik}{2\beta}+\frac{s+1}{2},\frac{1}{2},|\xi|^2\right)
    \mp \mathfrak{B} F\left(-\frac{ik}{2\beta}-\frac{s-1}{2},-\frac{ik}{2\beta}+\frac{s}{2}+1,\frac{3}{2},|\xi|^2\right) \notag\\
    &\equiv \mathfrak{A} F_A \mp \mathfrak{B} F_B,
\end{align}
where
\begin{align}
    \mathfrak{A} &= \frac{\Gamma\left(-\frac{ik}{\beta}+1\right)\Gamma\left(\frac{1}{2}\right)}
    {\Gamma\left(-\frac{ik}{2\beta}-\frac{s-1}{2}\right)\Gamma\left(-\frac{ik}{2\beta}+\frac{s}{2}+1\right)}, &
    \mathfrak{B} &= \frac{\Gamma\left(-\frac{ik}{\beta}+1\right)\Gamma\left(-\frac{1}{2}\right)}
    {\Gamma\left(-\frac{ik}{2\beta}-\frac{s}{2}\right)\Gamma\left(-\frac{ik}{2\beta}+\frac{s+1}{2}\right)} .
\end{align}
Substituting this into the quantization condition~\eqref{eq:quant_cond_F}, we find that
\begin{align}
    \left(\frac{\mathfrak{A} F_A + \mathfrak{B} F_B}{\mathfrak{A} F_A - \mathfrak{B} F_B}\right)^2 = 1 .
\end{align}
To obtain $(+1)^2$ on the left-hand side in Eq.~\eqref{eq:quant_cond_F}, it is easy to find $\mathfrak{B}=0$.
Therefore, $\Gamma\left(-\frac{ik}{2\beta}-\frac{s}{2}\right)$ or
$\Gamma\left(-\frac{ik}{2\beta}+\frac{s+1}{2}\right)$ should be at pole singularities.
Then, $-\frac{ik}{\beta}-s=-2n$ or $-\frac{ik}{\beta}+s+1=-2n$ ($n=0$, $1$, \dots).
The former corresponds to antiresonance and the latter is resonance,
being identical to the above observation in~Section~\ref{sec:rm_resonance} with \textit{even} numbers of nodes.
On the other hand, we also have another condition when the left-hand side in Eq.~\eqref{eq:quant_cond_F} becomes $(-1)^2$; then, $\Gamma\left(-\frac{ik}{2\beta}-\frac{s-1}{2}\right)$ or $\Gamma\left(-\frac{ik}{2\beta}+\frac{s}{2}+1\right)$ should diverge such that $\mathfrak{A}=0$.
Then, $-\frac{ik}{\beta}-s=-2n-1$ or $-\frac{ik}{\beta}+s+1=-2n-1$ ($n=0$, $1$, \dots).
This is a completely identical result to the complex energies with \textit{odd} numbers of nodes.
Totally, indeed, one finds that $-\frac{ik}{\beta}-s=-n$ or $-\frac{ik}{\beta}+s+1=-n$ ($n=0$, $1$, \dots), which is identical to~$E_n^{\rm{AR}}$ or $E_n^{\rm{R}}$, respectively.\footnote{In contrast to Gamow’s method which yields an exponentially small imaginary part for the energy from the leading tunneling action, the exact WKB method systematically incorporates all orders and treats the resonance condition with full analytical rigor (as in Eq.~\eqref{eq:quant_cond_F}).}

\section{Conclusion}\label{sec:conclusion}
In this paper, we have explored quasi-stationary quantum systems through the lens of the exact WKB analysis,
focusing on resonant states as a representative example of unstable phenomena.
By studying the inverted Rosen--Morse potential---a nonpolynomial potential exactly solvable system---we have shown how complex energies can be understood both from direct solution and from exact WKB quantization.
This approach provides a clear and analytically consistent picture of resonance, including barrier resonance.
We demonstrated the strength of exact WKB methods beyond conventional settings.

Our results suggest that the exact WKB framework offers a promising route to systematically understand nonperturbative aspects of resonance,
potentially bridging connections with other methods such as complex scaling and rigged Hilbert space formulations.
Further applications to more general systems would be a natural extension of this work.

\section*{Acknowledgements}
We would like to thank the organizers of Nambu Colloquium at Osaka University.
This work was partially supported by Japan Society for the Promotion of Science (JSPS)
Grant-in-Aid for Scientific Research Grant Numbers
JP22KJ2096 (O.M.) and JP21H04975 (S.O.).
O.M.\ acknowledges the RIKEN Special Postdoctoral Researcher Program.

\bibliographystyle{utphys}
\bibliography{ref,ref_ewkb}
\end{document}